\documentclass[%
 aip,
 amsmath,amssymb,
 reprint,%
]{revtex4-1}

\newcommand{\E}{{\mathbf e}}

\newcommand{\BEQ}{\begin{equation}} 
\newcommand{\EEQ}{\end{equation}} 
\newcommand{\BEA}{\begin{eqnarray}} 
\newcommand{\EEA}{\end{eqnarray}}

\newcommand{\PP}{\mathbb{P}}
\usepackage{graphicx}
\usepackage{dcolumn}
\usepackage{bm}

\usepackage[utf8]{inputenc}
\usepackage[T1]{fontenc}
\usepackage{mathptmx}
\usepackage{etoolbox}
\newtheorem{remark}{Remark}
\makeatletter
\def\@email#1#2{%
 \endgroup
 \patchcmd{\titleblock@produce}
  {\frontmatter@RRAPformat}
  {\frontmatter@RRAPformat{\produce@RRAP{*#1\href{mailto:#2}{#2}}}\frontmatter@RRAPformat}
  {}{}
}%
\makeatother
\begin{document}

\preprint{AIP/123-QED}

\title[Generalized time-fractional kinetic-type equations]{Generalized time-fractional kinetic-type equations with multiple  parameters}
\author{Luca Angelani}
\email{luca.angelani@cnr.it}
\affiliation{Istituto dei Sistemi Complessi - Consiglio Nazionale delle Ricerche, Piazzale A. Moro 5, I-00185, Roma, Italy
}%
\affiliation{Dipartimento di Fisica,
Sapienza Universit\`a di Roma, Piazzale A. Moro 5, I-00185, Roma, Italy
}%

\author{Alessandro De Gregorio}
\address{Department of Statistical Sciences, ``Sapienza" University of Rome, P.le Aldo Moro, 5 - 00185, Rome, Italy}
\email{alessandro.degregorio@uniroma1.it}

\author{Roberto Garra}
\address{Section of Mathematics, International Telematic University Uninettuno, Corso Vittorio Emanuele II, 39, 00186 Roma, Italy}
\email{roberto.garra@uninettunouniversity.net}

\date{\today}

\begin{abstract}
In this paper we study a new generalization of the kinetic equation emerging in run-and-tumble models. We show that this generalization leads to a wide class of generalized fractional kinetic (GFK) and telegraph-type equations depending by two (or three) parameters. 
We provide an explicit expression of the solution in the Laplace domain and   
show that, for a particular choice of the parameters, the fundamental solution of the GFK equation can be interpreted as the probability density function of a stochastic process obtained by a suitable transformation of the inverse of a subordinator. Then, we discuss some particular interesting cases, such as generalized telegraph models, diffusion fractional equations involving higher order time derivatives and fractional integral equations. 
\end{abstract}

\maketitle

\begin{quotation}
The interplay between fractional calculus, kinetic equations and stochastic subordination methods is, nowadays, a powerful tool to describe anomalous complex systems. Run-and-tumble models and their fractional generalizations represent  interesting attempts to study the displacements of particles scattering differently from classical Brownian diffusion. The main idea is to generalize the classical hyperbolic partial differential equations arsing in the analysis of the run-and-tumble motions, by means of some time-fractional Riemann-Liouville integrals and/or Caputo derivatives. This approach leads to a generalized version of the telegraph equation involving three parameters (or equivalently two parameters). This class of equations is wide and contains some interesting cases.  It is possible to determine the range of the parameters where the fundamental solution of the fractional telegraph equation is a probability density function. In this case, we are able to  interpret the stochastic solutions as anomalous diffusions by exploiting the theory of the Bernstein functions and the inverse subordinators. Furthermore, we study higher order fractional diffusion equations, where the fundamental solution can't be considered as a PDF of a random motion. Nevertheless, our equation is linked to heat diffusion involving higher order time derivatives. Fractional integral equations represent another non-trivial particular case, where the analysis of the asymptotic behaviour of the solution reveals that it turns back to the initial condition. This exotic behaviour should be clarified in future studies. 
\end{quotation}

\section{Introduction}

In the recent literature, fractional generalizations of the transport equation of active motion to describe the anomalous diffusion of self-propelled particles are gaining interest both for mathematical and physical reasons. For example, in Ref.\onlinecite{sevilla2023anomalous}, the authors studied the anomalous diffusion of self-propelled particles by using the fractional generalization of the transport equation of active motion. In Ref.\onlinecite{Ang_JSP2024}, a time-fractional generalization of the kinetic equation has been studied in the specific context of $d$-dimensional run-and-tumble models. These models are widely used for the study of finite velocity random motions of active particles, for example to describe the dynamics of motile bacteria 
Ref.\onlinecite{Ecoli_Berg,PhysRevE.48.2553}. 
The fractional generalization leads to non-trivial and peculiar characteristics. 
For example, the transport process obtained from the fractional kinetic equation is not still with finite velocity and has sample paths  trapped in some time intervals (see Ref.\onlinecite{Ang_JSP2024}). It is clear in this context the role played by the fractional generalization leading to a time-changed run-and-tumble model. The random motion related to the fractional generalization can be interpreted also in the framework of continuous-time random walks.
On the other hand, there is a wide literature about the connection between fractional telegraph equations and anomalous diffusion processes. For example in the recent paper Ref.\onlinecite{gorska24} the authors have shown that the fractional telegraph equation describe fractional-ballistic transport in the short-time regime, and transit to fractional diffusion transport in the long-time. Moreover, these models have been studied in the literature with different conditions in bounded domains Ref.\onlinecite{angelani2020fractional} and with stochastic resetting Ref.\onlinecite{gorska24}.

In this paper we start our analysis from the transport equation emerging in the context of the classical run-and-tumble models. Then, we introduce a new three-parameters generalization of the master kinetic equation by means of Riemann-Liouville fractional integrals. This mathematical generalization leads to a wide class of equations with different non-trivial behaviours depending on the choice of the parameters. We find the Laplace transform of the fundamental solution and the connection with fractional telegraph-type equations that are widely studied in the literature (see e.g. Ref.\onlinecite{CM1997,angelani2020fractional} and the references therein). Then, the central problem is to obtain the exact range of the parameters that guarantees that the solution is non-negative and can be interpreted as 
a probability density function (PDF) of a random motion. We obtain the answer to this problem by using the Bernstein's theorem (see Ref.\onlinecite{schilling-bern}) and the theory of the inverse subordinators (see Ref.\onlinecite{meerschaert2008triangular}). In this way we are also able to give a probabilistic interpretation of the related random motion. 

In the last part of the paper we discuss some interesting applications. First of all we consider the time-telegraph equation  clearly leading to an anomalous diffusion, as we prove by studying the 
mean-square displacement (MSD). Then, we consider the relation with heat equations involving higher order time derivatives and fractional integral equations. 
In conclusion, the generalized kinetic equation studied in this paper allows to obtain many different situations, some of which are quite peculiar and deserve consideration in further studies of higher order time-fractional diffusive equations.

\section{Generalized run-and-tumble model and the fractional kinetic equation}

The standard run-and-tumble model describes a particle that moves on $\mathbb R$  at constant speed $v$ and 
randomly reorients its direction of motion at rate $\alpha$ 
Ref.\onlinecite{weiss2002some,PhysRevE.48.2553}.
By denoting with  $P_{_R}(x,t)$ and  $P_{_L}(x,t)$ the PDF
of right-oriented and left-oriented  particles,
the kinetic equations describing the time evolution of the system are
(for simplicity we omit the explicit dependence on the variables $x$ and $t$)
\begin{eqnarray}
\label{eq_genR}
\frac{\partial P_{_R}}{\partial t} &= -& v \frac{\partial P_{_R} }{\partial x}
- \frac{\alpha}{2} (P_{_R} - P_{_L}) , \\
\label{eq_genL}
\frac{\partial P_{_L}}{\partial t} &= & v \frac{\partial P_{_L} }{\partial x}
+ \frac{\alpha}{2} (P_{_R} - P_{_L}) .
\end{eqnarray}
It is convenient, in view of a possible extension to the $d$-dimensional case 
Ref.\onlinecite{martens2012probability}, 
to rewrite the equations of the run-and-tumble model (RT) in a more concise form 
\begin{equation}
\label{RT}
\frac{\partial P_{\E}}{\partial t} = - v \E \frac{\partial P_{\E} }{\partial x}
+ \alpha  (\PP -1) P_{\E} \  , \qquad \E= \pm 1 \ , \qquad{\text{(RT)}}
\end{equation}
where $\E=\pm 1$ denotes the particle orientation with respect to the $x$-axis 
($P_{+1}\equiv P_{_R}$  and $P_{-1}\equiv P_{_L}$),
and we introduced the projector operator $\PP$
\begin{equation}
  \PP P_\E = \frac12 \sum_{\E=\pm 1} P_\E = \frac{P_{_R}+P_{_L}}{2}.
\end{equation}
Recently, a generalization of the kinetic run-and-tumble equation \eqref{RT}  has been studied in Ref.\onlinecite{Ang_JSP2024}, by considering
time-fractional derivatives in the sense of Caputo (for the theory of the fractional derivatives the reader can consult, e.g., Ref.\onlinecite{kilbas2006theory})
\begin{equation}
\label{CD}
    \frac{\partial^\nu f}{\partial t^\nu} (t) = \frac{1}{\Gamma(1-\nu)}
    \int_0^t d\tau \ (t-\tau)^{-\nu} \frac{\partial f}{\partial \tau} (\tau) \ ,
\end{equation}
with $\nu\in(0,1)$ and $\Gamma(\cdot)$ denotes the Euler gamma function.
In this case the fractional run-and-tumble equation (FRT) is obtained substituting the standard
time-derivative in (\ref{RT}) with the Caputo derivative (\ref{CD})
\begin{equation}
\label{FRT}
\frac{\partial^\nu P_{\E}}{\partial t^\nu} = - v \E \frac{\partial P_{\E} }{\partial x}
+ \alpha  (\PP -1) P_{\E} \  , \qquad \E= \pm 1 \ . \qquad {\text{(FRT)}}
\end{equation}
For an in-depth discussion of this equation and its interpretation see Ref. \onlinecite{Ang_JSP2024}.

In the spirit of previous studies on generalized fractional Cattaneo equations Ref.\onlinecite{CM1997,angelani2020fractional},
in this work we propose a broader fractional generalization of kinetic equations \eqref{RT} emerging in the run-and-tumble model.
We define the generalized fractional kinetic equations (GFK) as follows 
\begin{equation}
\label{GFRT}
I^{1-\nu} \Big( \frac{\partial P_\E}{\partial t}  \Big) =
- v \E \ I^{1-\mu} \Big( \frac{\partial P_\E}{\partial x}  \Big) +
\alpha(\PP-1) I^{1-\lambda} P_\E  ,
\end{equation}
with $\E= \pm 1 ,\nu,\mu,\lambda \in (0,1)$ and 
where $I^\nu$ is the Riemann-Liouville fractional integral (see Ref.\onlinecite{kilbas2006theory})
\begin{equation}
   I^{\nu} f(t) = \frac{1}{\Gamma(\nu)} \int_0^t d\tau (t-\tau)^{\nu-1} f(\tau) ,
       \qquad \nu>0 .
\end{equation}
We have introduced three  exponents, each associated 
to a fractional integral applied to the three terms of 
Eq. (\ref{RT}).
The GFK model is thus defined by
the triplet $\{\nu$, $\mu$, $\lambda\}$.
For $\mu=\lambda=1$, i.e., in the case $\{\nu$, $1$, $1\}$, 
the model reduces to the FRT (\ref{FRT}), investigated in \onlinecite{Ang_JSP2024}. 
Indeed, we have that  $\lim_{\nu \to 0} I^\nu f(t) = f(t)$ and 
the following identity holds (see \onlinecite{kilbas2006theory})
\begin{equation}
 \frac{\partial^\nu f}{\partial t^\nu} (t) =  I^{1- \nu} \frac{\partial f}{\partial t} (t) \ \quad \mbox{for $0<\nu<1$}.
\end{equation}
We also note that for $\nu=\mu=\lambda$ the equation (\ref{GFRT}) corresponds to the standard RT equation
(\ref{RT}), as $I^\nu f= 0$ has solution $f=0$. 

We solve the kinetic equation (\ref{GFRT}) in the Laplace-Fourier domain.
By considering the Laplace transform
\begin{equation}
    \tilde{f}(x,s) = \mathcal{L} [f(x,t)](x,s) = \int_0^\infty dt\ e^{-st} f(x,t) \ ,
\end{equation}
and the Fourier transform
\begin{equation}
    \hat{f}(k,t) = \mathcal{F} [f(x,t)](k,t) =\int_{-\infty}^\infty dx\ e^{ikx} f(x,t) \ ,
\end{equation}
the equation (\ref{GFRT}) becomes
\begin{equation}
    \Big( s^\nu +\alpha s^{\lambda-1} - i s^{\mu-1} v k \E \Big)
    {\hat{\tilde P}}_\E = s^{\nu-1} + \alpha s^{\lambda-1} \PP {\hat{\tilde P}}_\E \ ,
\end{equation}
where we have considered initial conditions
\begin{equation}
    P_\E(x,t\!=\!0) = \delta(x) \ ,
\end{equation}
and used the property (see, e.g., Ref.\onlinecite{kilbas2006theory})
\begin{equation}
    \mathcal{L} [I^{1-\nu} f(t) ] (s) = \frac{{\tilde f}(s)}{s^{1-\nu}} \ .
\end{equation}
By solving for ${\hat{\tilde P}}_\E $ and applying the operator $\PP$ 
we finally arrive at the expression of the Fourier-Laplace transform of the function  $P=\PP P_\E.$  Indeed, we have that
$${\hat{\tilde P}}(k,s)= \frac{s^{\nu-1}P_0(k,s)}{1-\alpha s^{\lambda -1}P_0(k,s)}$$
where
\begin{align*}
 P_0(k,s)&=\mathbb P\left(  \frac{1}{s^\nu+\alpha s^{\lambda-1}-i s^{\mu-1}vk \E}\right)\\
 &=\frac{s^\nu+\alpha s^{\lambda -1}}{(s^\nu+\alpha s^{\lambda-1})^2+ s^{2\mu-2}(vk)^2 }. 
\end{align*}
Then

\begin{equation}
\label{Pks}
    {\hat{\tilde P}}(k,s)= \frac{s^{\eta-1}(s^\epsilon + \alpha)}{s^{\eta}(s^\epsilon +\alpha)+ v^2 k^2} \ ,
\end{equation}
where the exponent $\eta$ and $\epsilon$ are 
\begin{align}
\label{eta}
    \eta &=\eta(\nu,\lambda,\mu)= 1 +\nu+\lambda -2 \mu \ , \\
    \epsilon &=\epsilon(\nu,\lambda)= 1+\nu-\lambda \ .
\label{eps}
\end{align}
We should underline that we have used the notation of the classical kinetic theory, but in this case a non-trivial problem is to understand the cases in which the function $P$ can be properly considered as a probability density of a random process. 
In the following section we can give an answer to this problem.
On the other side, with this generalization of the kinetic equation we recover as particular cases diffusive models involving higher order time derivatives. This is particularly interesting, since in the literature there are few studies about fractional (or non fractional) diffusive models involving time derivatives of order greater that two. 

Starting from \eqref{Pks} we can easily show that the function ${\hat{\tilde P}}(k,s)$ coincides with the 
Fourier-Laplace transform of the solution of a time-fractional telegraph-type equation. 
Indeed, from \eqref{Pks} we have that
\begin{equation}
s^{\eta+\epsilon}{\hat{\tilde P}}(k,s)+\alpha s^\eta {\hat{\tilde P}}(k,s)- s^{\eta+\epsilon-1}-\alpha s^{\eta-1} = -v^2k^2{\hat{\tilde P}}(k,s)
\end{equation}
and the corresponding fractional equation is given by 
\begin{equation}\label{tel}
 \frac{\partial^{\eta+\epsilon}P}{\partial t^{\eta+\epsilon}}  +\alpha \frac{\partial^{\eta}P}{\partial t^{\eta}} = v^2 \frac{\partial^2 P}{\partial x^2},
\end{equation}
with $P=P(x,t),$
and the initial conditions are given by
$$
 P(x,0) = \delta(x)$$
 if $0<\eta +\epsilon \leq1$,
    $$P(x,0) = \delta (x), \, \partial_t P\big|_{t=0}=0
    $$
    if $1<\eta +\epsilon\leq2$,
    $$P(x,0) = \delta (x), \quad \partial_t P\big|_{t=0}=\partial^2_t P\big|_{t=0} =0$$ if $2<\eta +\epsilon \leq 3$,
    $$P(x,0) = \delta (x), \, \partial_t P\big|_{t=0}=\partial^2_t P\big|_{t=0} =\partial^3_t P\big|_{t=0} =0$$ if $3<\eta +\epsilon \leq 4$.
Indeed, we underline that the formal derivation of the governing equation \eqref{tel} leads
to a time-fractional equation depending by the parameters $\eta$ and $\epsilon$, whose range in this case is not restricted to $(0,1)$, as in the classical fractional telegraph equation. 
This is one of the most interesting and new point related to this generalization. Indeed, by exploiting a reasonable generalization, we obtain a higher order fractional kinetic-type equation. Moreover, our analysis started from a three parameters generalization (namely $\nu$, $\lambda$ and $\mu$), but we have shown that the governing equation is reduced to a two parameter telegraph-type equation of order $\eta$ and $\epsilon$. This is an interesting object for the mathematical analysis. We underline that in the recent paper Ref.\onlinecite{Enzo} the authors considered higher order fractional equations  starting from the generalization of a probabilistic model, obtaining interesting results.\\
Observe that, under suitable conditions (for details we refer e.g. to Ref.\onlinecite{Ang_JSP2024}), the fractional equation \eqref{tel}  corresponds to the set of equations
\begin{align}
\nonumber & \displaystyle\frac{\partial^{\eta}P}{\partial t^{\eta}}  = - \frac{\partial J}{\partial x},\\
\nonumber & \displaystyle\frac{\partial^{\epsilon}J}{\partial t^{\epsilon}}+ \alpha J  = - v^2\frac{\partial P}{\partial x},
\end{align}
that suggests a possible interpretation in the context of the studies about fractional heat equations and anomalous diffusion models (see, e.g., Ref.\onlinecite{CM1997}). However, also in this case, this is a formal analogy in view of the wide range of the parameters from the general approach. 

The inverse Fourier transform of (\ref{Pks}) gives
the explicit solution in the $(x,s)$ domain
\begin{equation}
\label{Pxs}
    \tilde{P}(x,s) = \frac{\sqrt{s^\eta(s^\epsilon+\alpha)}}{2vs}
 \ \exp{\Big( - \frac{\sqrt{s^\eta(s^\epsilon+\alpha)}}{v} |x| \Big)} \ .
 \end{equation}
We emphasize again that 
the above expression 
depends on the exponents $\{\nu,\mu,\lambda\}$ only thorough the two exponents 
$\{\eta,\epsilon\}$, whose range of values is
$\eta \in (-1,3)$ and $\epsilon \in (0,2)$ with $0<\epsilon+\eta<4$ and $-2<\epsilon-\eta<2$ 
(see Figure \ref{fig1}).
However, not all  values correspond to a non-negative 
 function $P(x,t)$ and, therefore,  
 there are regions in the parameters space in which the function $P(x,t)$
 cannot  be interpret as a PDF.  
\section{Stochastic solutions to  the generalized telegraph-type equation}

We now study the conditions that must be satisfied by the parameters $\{\eta,\epsilon\}$ in order to interpret $P(x,t)$ 
as the PDF of a random motion. 

 It is useful to recall the following definitions (see Ref.\onlinecite{schilling-bern}). A function $f:(0,\infty) \to \mathbb R$ is a completely monotone function (CM) if $f$
is of class $C^\infty$ and $(-1)^{n} f^{(n)}(s)>0$ for all $s>0$ and non-negative integer $n$. The function $f$  is called Bernstein function (BF) if $f(s)\geq 0,$ for all $ s\geq 0,$ and  $f$
is of class $C^\infty$ and $(-1)^{n-1} f^{(n)}(s)>0$ for all $s>0$ and any $n\in\mathbb N$. We recall that $f(s)=s^\lambda$ is a Bernstein function if and only if $0\leq \lambda \leq 1.$ Furthermore, the Bernstein function $f$ admits the Lévy-Khintchine representation 
$$f(s)=a+bs+\int_{(0,\infty)}(1-e^{-st})\mu(dt)$$
where $a, b \geq 0$ and $\mu$ is a measure on $(0,\infty)$ satisfying $\int_{(0,\infty)}(1\wedge t)\mu(dt)<1$.

A function $f$ is called Complete Bernstein function (CBF) if $f$ is a Bernstein function and its Lévy measure has a completely monotone density $m(t)$ with respect to Lebesgue measure such that the Lévy-Khintchine representation becomes
$$f(s)=a+bs+\int_{(0,\infty)}(1-e^{-st})m(t)dt.$$
The function $f(s)=s^\lambda$ is also a Complete Bernstein function since $m(t)=\frac{\lambda}{\Gamma(1-\lambda)t^{\lambda +1}},$ with $0< \lambda < 1,$ and $a=b=0.$

 Let us observe that the function $$z(s)=\sqrt{s^\eta(s^\epsilon+\alpha)}$$
is CBF for $0<\eta<1$ and $0<\epsilon<1.$ Indeed, by exploting Propostion 7.13 in Ref.\onlinecite{schilling-bern}, we have that $s^\eta$ and $s^\epsilon+\alpha$ are CBFs for $0<\eta<1$ and $0<\epsilon<1,$ respectively, and then $z(s)$ is CBF as well.

\begin{figure}[t!]
\includegraphics[width=1\linewidth] {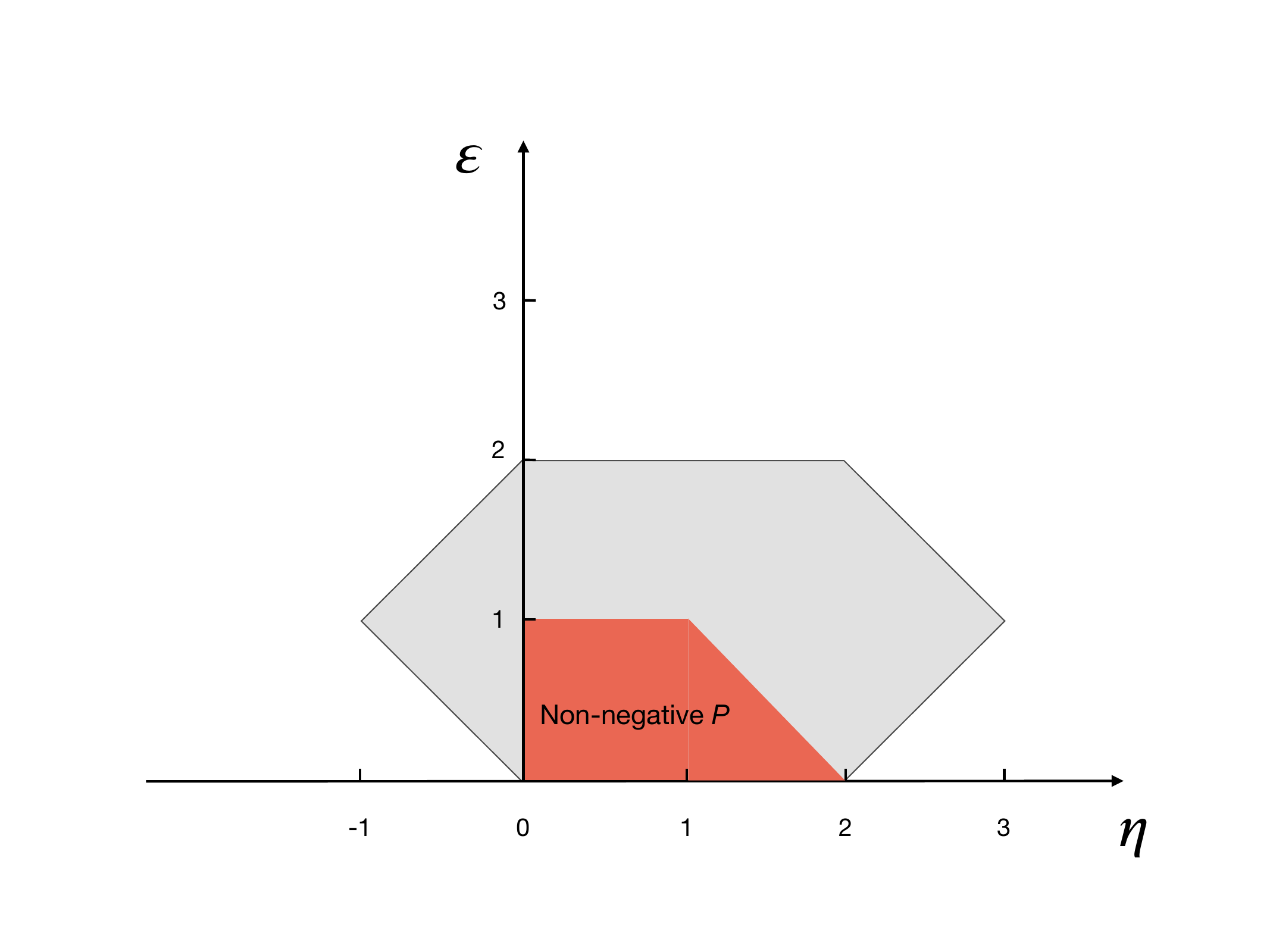}
\caption{\label{fig1}
Range of values of exponents $(\eta,\epsilon)$ of the model, see (\ref{Pks}), (\ref{tel}) and (\ref{Pxs}).
The coloured area represents the region $\{(\eta,\epsilon) : 0<\eta<2 , 0<\epsilon< \min(1,2-\eta)\}$
where the function $P(x,t)$ can be interpreted as a PDF of a random motion.
}
\end{figure}

We now show that the function $z(s)$ is CBF even for $1<\eta<2$ when $0<\epsilon<2-\eta$.
Indeed, we can write, for a given $\xi \in(0,2)$,
$$
z(s) = f(s)^{\xi/2} g(s)^{1-\xi/2} ,
$$
where $f(s)=s^{\eta/\xi}$ and $g(s)=(s^\epsilon +\alpha)^{\frac{1}{2-\xi}}$.
The function $f$ is CBF for $0<\eta<\xi$. 
The function $g$ can be written as $g(s) = h(s^\epsilon)^{1/\epsilon}$ 
where $h(s)=(s+\alpha)^{\frac{\epsilon}{2-\xi}}$.
Now we observe that $h(s)$ is CBF for $0<\epsilon<2-\xi$, and then, from Corollary 7.15 in Ref.\onlinecite{schilling-bern},
we have that, for $0<\epsilon<1$, $g(s)$ is CBF as well.
For a given $\xi \in (0,2)$ we can exploit, as before, Proposition 7.13 in \onlinecite{schilling-bern} 
and conclude that $z(s)$ is CBF for $0<\eta<\xi$ and $0<\epsilon<\min(1,2-\xi)$ (the upper limit 
is given by the more stringent condition on $\epsilon$ for $g$ to be CBF).
For $\xi\in(0,1)$ we obtain the previous result that $z(s)$ is CBF for for $0<\eta<1$ and $0<\epsilon<1$.
By considering $\xi\in(1,2)$ we now conclude that $z$ is CBF even in the region $1<\eta<2$ and $0<\epsilon<2-\eta$
(see Figure \ref{fig1}).\\
In the region where $z(s)$ is CBF, the function ${\tilde P}$ (\ref{Pxs}) 
is CM. Indeed, we have 
$$\tilde P(x,s) = \frac{z(s)}{2vs} e^{-|x| z(s) /v} ,$$
and we use the following properties
(we remind that CBF is a subclass of BF):
if $f(s)$ is BF then  $e^{-a f(s)}$ (with $a>0$) and 
$f(s)/s$ are CM
(see Theorem 3.6 and Corollary 3.7 in \onlinecite{schilling-bern})
;
if $f(s)$ and $g(s)$ are CM also their product $f(s) g(s)$ is CM
(Corollary 1.6).
By  Bernstein's theorem Ref.\onlinecite{schilling-bern}, we have that a function is  completely monotone 
if and only if it is the Laplace transform of a non-negative measure.
Then we conclude that the function $P(x,t)$ is non-negative in the parameter region
$\{(\eta,\epsilon) : 0<\eta<2 , 0<\epsilon< \min(1,2-\eta)\}$
(colored area in Figure \ref{fig1}).
It is worth noting that the above arguments are valid for $\alpha \geq 0$, however
for $\alpha=0$  there are less stringent conditions for the non-negativity of $P$.
Indeed, in such a case, the function $z(s)=s^{\frac{\eta+\epsilon}{2}}$ is CBF for 
$0<\eta+\epsilon<2$ and, therefore, $P$ is non-negative in a wider area of the parameter space.

 Now, in the region where $z(s)$ is CBF, we are able to provide explicitly $P(x,t)$ and give a clear interpretation of the stochastic process governed by the fractional telegraph equation \eqref{Pxs}.  Since $z(s)$ is a CBF, there exists a subordinator $L:=\{L(t):t\geq 0\},$ that is a non-decreasing and non-negative Lévy process, with Laplace exponent $z(s)$ (see Proposition 1.3.24, Ref.\onlinecite{applebaum2009levy}); i.e. 
$$\mathbb E(e^{-s L(t)})=e^{-tz(s)}.$$
For the definition of subordinator the reader can consult Ref.\onlinecite{applebaum2009levy}.
Let $Y(t)$ be the inverse of $L(t)$ at time $t\geq 0;$ i.e. $Y(t):=\inf\{u>0: L(u)>t\}.$ We observe that $P(Y(t)\leq x)=P(L(x)\geq t),$ for any $x\geq 0.$  The subordinator $L$ is driftless since in $\{(\eta,\epsilon) : 0<\eta<2 , 0<\epsilon< \min(1,2-\eta)\},$ we have that $$\lim_{s\to\infty}\frac{z(s)}{s}=0,$$
and then $z(s)=\int_{(0,\infty)}(1-e^{-st})m_L(t)dt,$ where $m_L(t)dt$ is the Lévy measure of $L.$
Furthermore, $m_L(u,\infty):=\int_u^\infty m_L(t)dt$ is unbounded, that is $m(0,\infty)=\infty,$ because $\lim_{s\to\infty}z(s)=\infty.$  
Therefore, $L$ is strictly increasing and by applying Theorem 3.1 in Ref.\onlinecite{meerschaert2008triangular}, we can conclude that $Y(t)$ admits Lebesgue density given by $w(x,t)=\int_0^tm_L(t-u,\infty) P_{L(x)}(du),$ where $P_{L(x)}(du) $ is the probability measure of $L(x).$ For other smoothness properties on the density of the inverse of a subordinator the reader can consult Ref.\onlinecite{ascione2023regularity}. Furthermore, we observe that  $$w(x,t)=-\frac{d}{dx}P(L(x)\leq t)=-\frac{d}{dx}\int_0^t P_{L(x)}(du).
$$
 Given $x\geq 0,$ we can write down
\begin{align*}
\int_0^\infty e^{-st} w(x,t)dt &=-\int_0^\infty e^{-st}\left(\frac{d}{dx}\int_0^t  P_{L(x)}(du)\right) dt\\
&=-\frac1s\frac{d}{dx}\int_0^\infty  e^{-s u} P_{L(x)}(du) \\
&=-\frac1s\frac{d}{dx}\mathbb E(e^{-s L(x)})\\
&=\frac{z(s)}{s}e^{-x z(s)},
\end{align*}
in agreement with the result (3.13) in Ref.\onlinecite{meerschaert2008triangular}.
Let us define $$Z(t):= v D Y(t),$$ for any $t\geq0,$ where $D$ is a random variable with $P(D=1)=P(D=-1)=\frac12,$ independent from $Y(t).$ It is not hard to prove that the density function of $Z(t)$ is equal to $\frac{1}{2v}w(\frac{|x|}{v},t).$ Finally, we can conclude that $\frac{1}{2v}w(\frac{|x|}{v},t)$ coincides with $P(x,t)$ since
$$\frac{1}{2v} \int_0^\infty e^{-st} w\left(\frac{|x|}{v},t\right)dt =\tilde P(x,s). $$

\begin{remark}
It is worth mentioning that if $\alpha=0$ and $0<\eta+\epsilon<1,$ from \eqref{tel} we obtain that $P(x,t)$ becomes the fundamental solution of the time-fractional diffusion equation
\begin{equation}\label{eq:tfde}
\frac{\partial^{\eta+\epsilon}P}{\partial t^{\eta+\epsilon}}=v^2\frac{\partial^2 P}{\partial x^2},
\end{equation}
with $P(x,0)=\delta(x).$ Furthermore, in this setting $L(x),x\geq 0,$ is the stable subordinator with Laplace exponent $z(s)=s^{\frac{\eta+\epsilon}{2}}$ and measure $$m_L(dt)=\frac{\frac{\eta+\epsilon}{2}}{\Gamma(1-\frac{\eta+\epsilon}{2})t^{\frac{\eta+\epsilon}{2} +1}}dt.$$
Therefore, we can conclude that $P(x,t)$ is the PDF of the delayed Brownian motion $B(Y(t)), t\geq 0,$ where $B(t),t\geq0,$ is a Brownian motion and $Y(t)$ is the inverse of $L(x),x\geq 0$ (see, e.g., Ref.\onlinecite{meerschaert2019inverse}).

Let us consider $1<\eta+\epsilon<2;$ in this case the fractional wave equation \eqref{eq:tfde} still admits stochastic solutions. For a discussion of this topic the reader can consult Ref.\onlinecite{meerschaert2015stochastic} and Ref.\onlinecite{fujita1990cauchy}.
\end{remark}

\begin{remark}
If $\alpha>0$ and $\eta=\epsilon,$ with $\frac12<\eta<1,$ the equation \eqref{tel} reduces to the  following Cauchy problem involving a fractional telegraph equation (with one parameter)
\begin{align}\label{eq:cte}
&\frac{\partial^{2\eta}P}{\partial t^{2\eta}}+\alpha\frac{\partial^{\eta}P}{\partial t^{\eta}}=v^2\frac{\partial^2 P}{\partial x^2},\\
& P(x,0)=0,\partial_t P(x,t)|_{t=0}=0,\notag
\end{align}
studied by several authors (see, e.g., Ref.\onlinecite{orsingher2004time}, \onlinecite{CM1997}). In Ref.\onlinecite{Ang_JSP2024}, the stochastic solution of \eqref{eq:cte} is interpreted as the PDF of the time-changed telegraph process $X(Y(t)),t\geq 0,$ where $X(t)=D\int_0^t (-1)^{N(s)}ds$ is the telegraph process ($N(t)$ is a Poisson process with rate $\alpha/2$ independent from $D$) and $Y(t)$ is the inverse of a stable subordinator with $z(s)=s^{\eta}.$
\end{remark}

\section{Particular interesting cases}

\subsection{The generalized telegraph equation}

In the case $0<\eta<1$ and $0<\epsilon<1$
we recover a time fractional telegraph equation with two parameters, widely studied in the physical and probabilistic literature (see, e.g., Ref.\onlinecite{angelani2020fractional,CM1997,orsingher2004time,Ang_JSP2024} and the references therein). 
We underline that  most  studies in the literature are devoted to the fractional telegraph equation dependent on a single parameter. The present study, not only introduce a second parameter, but also
allows us to extend the region of investigation in the parameter space.
Indeed, as elucidated in previous sections, we have demonstrated that  
the function $P(x,t)$ can be interpreted as a PDF of a random motion 
in the extended zone 
$\{(\eta,\epsilon) : 0<\eta<2 , 0<\epsilon< \min(1,2-\eta)\}$.
The obtained fractional telegraph equation is related to an anomalous diffusion model, as we are going to prove by studying the MSD.\\
Indeed, the MSD can be obtained from the second derivative of the Fourier transformed PDF
\begin{equation}
  r^2 (t) = - \left. \frac{\partial^2 {\hat P}}{\partial k^2} 
       \right|_{k=0} ,
\end{equation}
or, in the Laplace domain
\begin{equation}
    {\tilde {r^2}} (s) = - \left. \frac{\partial^2 {\hat {\tilde P}}}{\partial k^2} 
       \right|_{k=0} .
\end{equation}
Using (\ref{Pks}) we  obtain  
\begin{equation}
\label{r2s}
    {\tilde {r^2}} (s) = \frac{2 v^2}{s^{\eta+1}(s^\epsilon+\alpha)} \ ,
\end{equation}
whose inverse Laplace transform can be written in term of the two-parameters Mittag-Leffler fucntion
$E_{\mu,\nu}$  Ref.\onlinecite{gorenflo2020mittag}
\begin{equation}
    E_{\mu,\nu} (z) = \sum_{n=0}^\infty \frac{z^n}{\Gamma(n\mu+\nu)} \ , 
    \qquad z \in {\mathbb C}, \ \mu,\nu>0 \ .
\end{equation}
Indeed, by using the property
\begin{equation}
    \mathcal{L}[t^{\nu-1} E_{\mu,\nu}(a t^\mu)](s) = \frac{s^{\mu-\nu}}{s^\mu-a} 
    \qquad \text{Re}\ \mu,\nu>0 \ ,
\end{equation}
we finally obtain the expression of the MSD
\begin{equation}
    \label{r2t}
    r^2(t) = 2 v^2 t^{\eta+\epsilon} \ E_{\epsilon,1+\eta+\epsilon}(-\alpha t^\epsilon) \ .
\end{equation}
By using Tauberian theorems Ref.\onlinecite{klafter2011first}, the long time regime of the MSD can be obtained from the 
behavior at small $s$ of (\ref{r2s}).
For $s \to 0$ we have
\begin{equation}
{\tilde{r^2}}(s) \sim \frac{2v^2}{\alpha s^{\eta+1}} \  ,
\end{equation}
corresponding in the time domain to
\begin{equation}
    r^2(t) \sim  
    \frac{2v^2}{\alpha \Gamma(\eta+1)} \ t^{\eta} \ .
       \hspace{1.3cm} t \to \infty .
\end{equation}
Therefore, if $\eta \in (0,1)$ we have a subdiffusion, while for $\eta\in(1,2)$ we have a superdiffusion, independently from the value of $\epsilon$.
At short times the dominant term is obtained for $s \to \infty$
\begin{equation}
{\tilde{r^2}}(s) \sim   \frac{2v^2}{s^{1+\eta+\epsilon}} \ ,
\end{equation}
corresponding to 
\begin{equation}
    r^2(t) \sim 
    \frac{2v^2}{\Gamma(1+\eta+\epsilon)} \ t^{\eta+\epsilon}\  ,
       \hspace{1.3cm} t \to 0 .
\end{equation}

{\subsection{Diffusive equations involving higher order time derivatives}
We observe that the fractional equation \eqref{tel} includes as special cases also  models involving time derivatives greater than two. For example, we can consider the case $\alpha = 0$ and $\eta+\epsilon = 3$ leading to the equation
\begin{equation}\label{3p}
\frac{\partial^3 P}{\partial t^3} = v^2\frac{\partial^2 P}{\partial x^2} .
\end{equation}
By considering initial conditions
\begin{align}
 P(x,0) &=\delta(x) , \\
 \partial_t P(x,t) \big|_{t=0} &=0  , \\
 \partial^2_t P(x,t) \big|_{t=0} &=0  ,
\end{align}
the solution of (\ref{3p}) in the Laplace-Fourier space is
\begin{equation}
\label{3pks}
    {\hat{\tilde P}}(k,s)=\frac{s^2}{s^3 + v^2 k^2} \ ,
\end{equation}
as can be easily obtained by direct calculation or using (\ref{Pks})
with our choice of parameters.
The inverse Fourier transform reads
\begin{equation}
    \tilde{P}(x,s) = \frac{s^{1/2}}{2v} \ \exp\left( -\frac{|x|}{v} s^{3/2}\right) .
\end{equation}
The above function is clearly non-monotone and, therefore, the $P(x,t)$ function cannot be considered as a probability
function of a random motion.
However, we can establish a link with diffusive models involving higher order time derivatives 
considered in the literature, for example, by Monin (we refer to the review paper Ref.\onlinecite{bakunin}, Section 9). 
Indeed, we can write (\ref{3pks}) as a sum of two terms
\begin{equation}
        {\hat{\tilde P}}(k,s)=\frac13 \left[ {\hat{\tilde P}}_1(k,s) +  {\hat{\tilde P}}_2(k,s)  \right] ,
\end{equation}
where 
\begin{align}
 {\hat{\tilde P}}_1(k,s) &= \frac{1}{s+(v^2k^2)^{1/3}} , \\
 {\hat{\tilde P}}_2(k,s) &= \frac{2s-(v^2 k^2)^{1/3}}{s^2+(v^2 k^2)^{2/3} - s (v^2 k^2)^{1/3}} .
\end{align}
It is interesting to note that the first function, in the $(k,t)$ domain, reads
\begin{equation}
    \hat{P}_1(k,t) = e^{-|v k|^{2/3}t} ,
\end{equation}
which corresponds to the Fourier transform of a $\frac23$-stable process
solution of (\ref{3p}).
Also, it is remarkable that it coincides with the Fourier transform of the solution of the space-fractional equation 
\begin{equation}\label{3pf}
\frac{\partial P}{\partial t} = v^2\frac{\partial^{2/3} P}{\partial |x|^{2/3}},
\end{equation}
under suitable conditions. This is a non-trivial relation that can be useful to give an interpretation to the anomalous diffusion models governed by equations of the form \eqref{3pf} or, more generally, involving fractional derivatives higher than two. \\
Moreover, diffusive equations involving higher order time derivatives emerges in the studies about heat transfer in metal films. We refer in particular to the recent paper Ref.\onlinecite{Ignaczak}. 
In this paper the author considered a generalization of the Cattaneo law that is given by
\begin{equation}\label{ca}
\left(1+\tau_q\frac{\partial}{\partial t}+\frac{1}{2}\tau_q^2\frac{\partial^2}{\partial t^2}\right)q(x,t) = -k\left(1+\tau_T\frac{\partial}{\partial t}\right) T(x,t),
\end{equation}
where $q(x,t)$ is the heat flux, $T(x,t)$ is the temperature field, $\tau_q$ and $\tau_T$ denote the phase lag of the heat flux and the phase lag of the temperature gradient, respectively.
The generalized Cattaneo law \eqref{ca} coupled with the energy equation
\begin{equation}
\rho c\frac{\partial T}{\partial t} =  - \frac{\partial q}{\partial x},
\end{equation}
leads to a model equation of heat diffusion involving third order time derivatives. A time-fractional generalization of these equations can be an interesting object of research in view of the wide applications of fractional calculus in anomalous diffusive models and fractional Cattaneo law.
Our derived fractional kinetic equation \eqref{tel} is strictly related to these models and can be obtained by a similar derivation. However a complete study about the fractional generalization of this model should be object of future research. 

\subsection{The fractional integral equation ($\eta<0$)}
We now briefly consider the case of negative exponent $\eta \in (-1,0)$. Let be $\beta = -\eta$, in this case 
the GFK model corresponds to the following equation (with $\epsilon>\beta$)
\begin{equation}
\label{gfi}
   \frac{\partial^{\epsilon-\beta}P}{\partial t^{\epsilon-\beta}}  +
   \alpha I^{1+\beta} \frac{\partial P}{\partial t} = v^2 \frac{\partial^2 P}{\partial x^2} ,
\end{equation}
with
$$
I^{1+\beta} \frac{\partial P}{\partial t} = I^\beta P - P|_{t=0} \frac{t^\beta}{\Gamma(1+\beta)},
$$
where $P|_{t=0}  = P(x, t=0) = \delta(x)$. 

In the Laplace-Fourier domain the solution of (\ref{gfi}) turns out to be
\begin{equation}
\label{Pksn}
    {\hat{\tilde P}}(k,s)=\frac{1}{s} \ \frac{s^\epsilon + \alpha}{s^\epsilon +\alpha+ s^\beta v^2 k^2} \ ,
\end{equation}
corresponding, in the $(x,s)$ domain, to
\begin{equation}
\label{Pxsn}
    \tilde{P}(x,s) = \frac{1}{2vs}\sqrt{\frac{s^\epsilon+\alpha}{s^\beta}}
    \ \exp{\left( -\frac{|x|}{v} \ \sqrt{\frac{s^\epsilon+\alpha}{s^\beta}}  \right)} \ .
 \end{equation}
Two remarks are in order, valid for $\alpha>0$.
First we note that the function (\ref{Pxsn}) is non-monotone, as it is immediately evident
from the fact that it goes to zero both for $s\to 0$ and $s\to \infty$ 
(we remind that $0<\beta < \epsilon$). Therefore, the solution
$P(x,t)$ is not non-negative defined and it can not be interpreted as a PDF of a random motion.
Second, we infer an unusual behavior of the function $P(x,t)$. Indeed, by studying the asymptotic limit
$t\to\infty$, we have
\begin{equation}
{\hat P}(k,t \to \infty) = \lim_{s \to 0} s {\hat{\tilde P}}(k,s) =
\lim_{s \to 0} \frac{s^\epsilon+\alpha}{s^\epsilon+\alpha+ s^\beta v^2k^2} = 1 \ ,
\end{equation}
corresponding to a solution that, asymptotically, turns back to the initial condition
\begin{equation}
    P(x,t \to \infty) = \delta (x) \ .
\end{equation}
The meaning and the possible relevance of this model for the description of physical processes deserve further 
investigation.

\section{Conclusions}

In this paper we have studied a time-fractional generalization of the kinetic equation emerging in the context of run-and-tumble models by introducing three different parameters (the order of the fractional integrals appearing in the master equation (7)). 
We have shown that the GFK equation can be reduced to a more simple fractional telegraph-type equation with two independent parameters. The main novelty is given by the wide class of models and equations related to this generalization, depending on the choice of  parameters. We have found the conditions that guarantee the non-negativity of the solution and a probabilistic interpretation, also in connection with the time-fractional telegraph equation with two independent parameters. However, there are also other peculiar cases that in our view should be object of further studies. 
For example, for a given choice of parameters, we obtain higher order fractional diffusion equations. In this case the fundamental solution can't be considered as a PDF of a random motion but it is interesting for mathematical reasons and in relation to heat diffusion involving higher order time derivatives. 
Another interesting case is $\eta<0$, since, from the analysis of the asymptotic behaviour we have a solution that turns back to the initial condition, this is a non-trivial and exotic behaviour that needs to be clarified in future studies. 



\begin{acknowledgments}
The authors would like to thank Prof. Bruno Toaldo (University of Turin - Italy) for the helpful discussion about some points of the paper. LA acknowledges financial support from the Italian Ministry of University and Research (MUR) 
under  PRIN2020 Grant No. 2020PFCXPE. The research of ADG is partially supported by  Italian Ministry of University and Research (MUR) under PRIN 2022 (APRIDACAS), Anomalous Phenomena on Regular and Irregular Domains: Approximating Complexity for the Applied Sciences, Funded by EU - Next Generation EU
CUP B53D23009540006 - Grant Code 2022XZSAFN - PNRR M4.C2.1.1.
\end{acknowledgments}

\bibliography{references}

\end{document}